\documentclass[twocolumn,aps,prl,superscriptaddress,amsfonts]{revtex4-1}

\usepackage{amsmath}
\usepackage{graphicx}
\usepackage{bm}
\usepackage{array}
\usepackage{dcolumn}
\usepackage{hepparticles}
\usepackage{heppennames}
\usepackage{hepnicenames}
\usepackage{hyperref}

\newcommand{\FigureOne}{
\begin{figure}[htbp!]
\includegraphics*[width=\columnwidth]{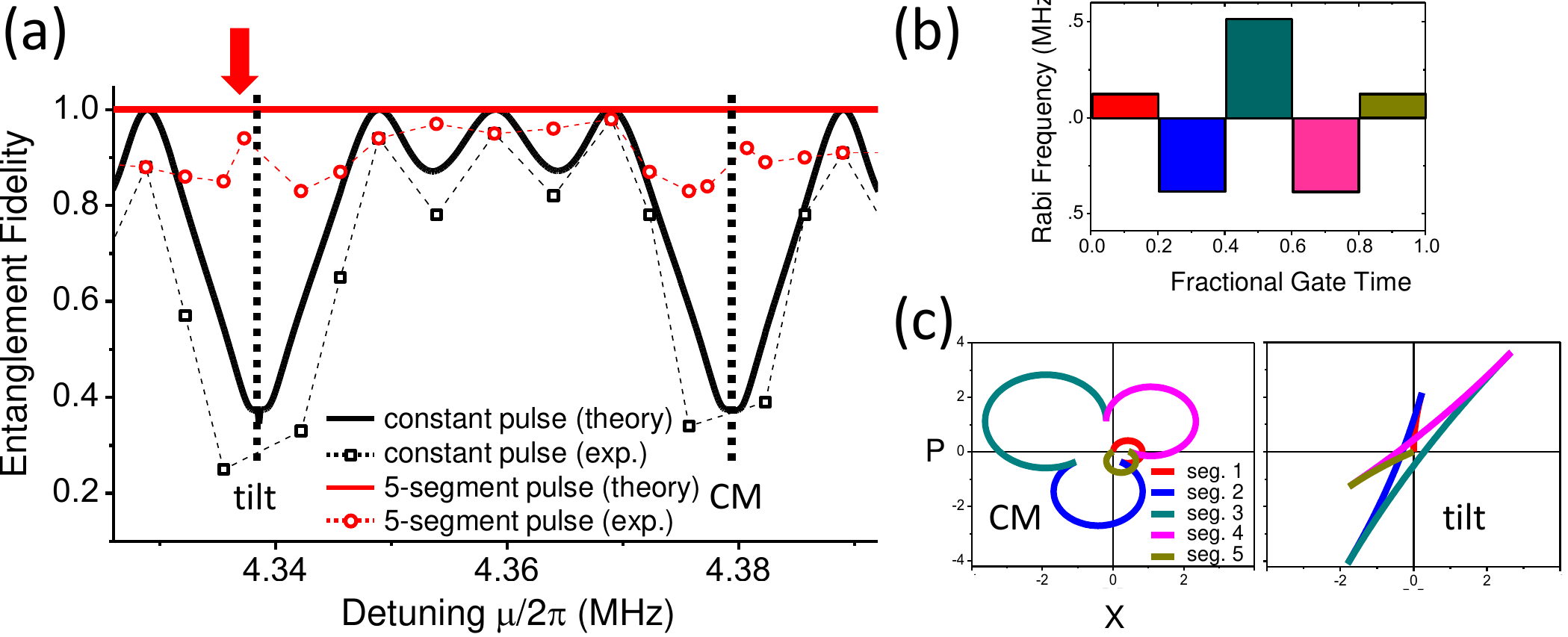}
\caption{Improvement of entangled state creation using pulse shaping on $N=2$ trapped ion qubits. 
(a) Comparison of Bell state entanglement fidelity for a constant pulse (black) versus a five-segment pulse (red) over a range of detuning $\mu$, showing significant improvement with the segmented gate. 
(b) The segmented pulse pattern, parameterized by the Rabi frequency $\Omega_i(t)$ with the particular detuning $\mu$ 
near the $2^{nd}$ (``tilt") motional mode (arrow in (a)) and measured state fidelity $\geq$94(2)\%. 
(c)  Phase space trajectories (arbitrary units) subject to pulse sequence in (b) for both CM and tilt modes of the two ions. 
The five-segment pulse pattern brings the two trajectories back to their origins, simultaneously disentangling both modes of motion from the qubits.}
\label{fig:fullControl}
\end{figure}}

\newcommand{\FigureTwo}{
\begin{figure}[htbp!]
\includegraphics*[width=\columnwidth]{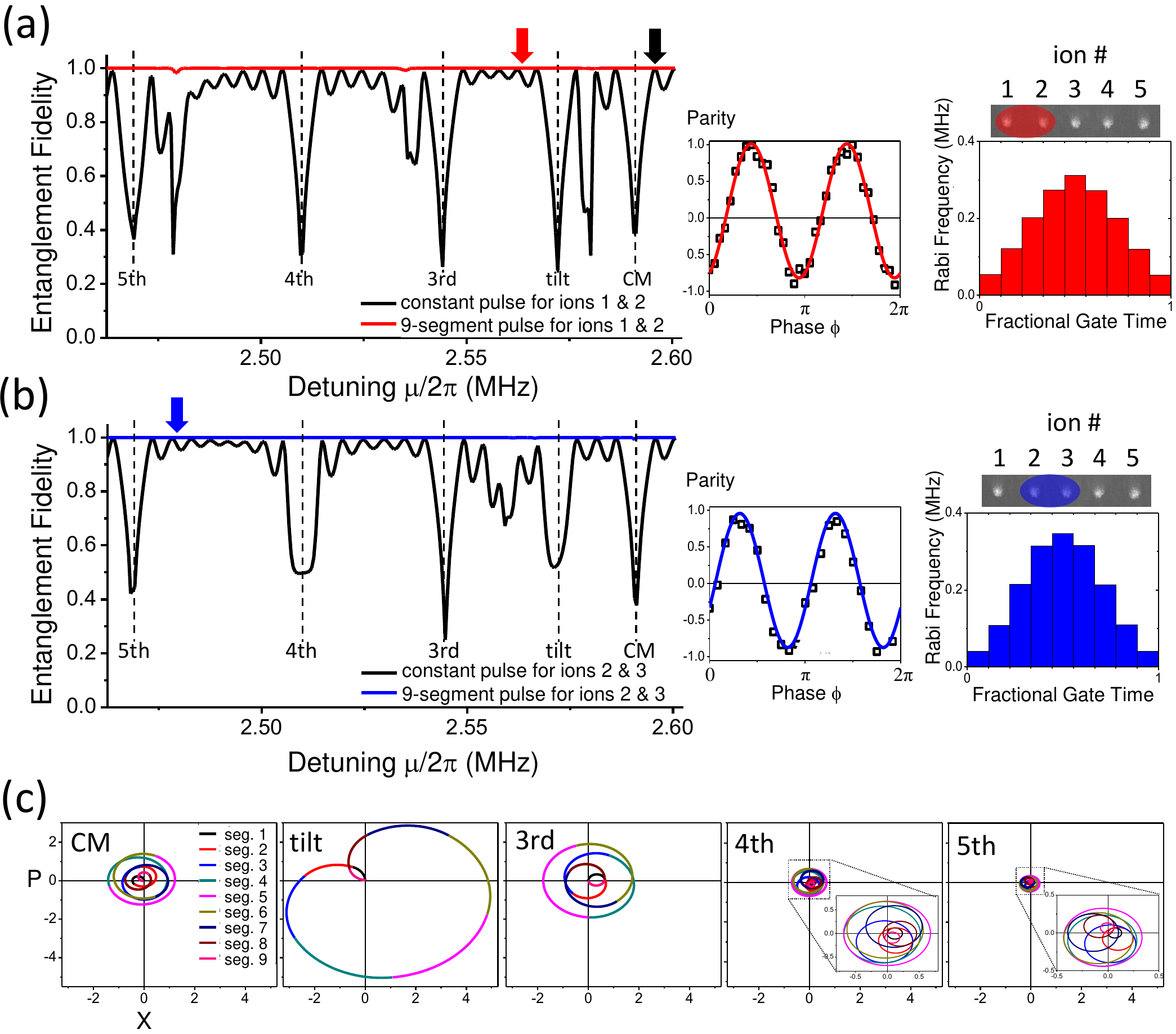}
\caption{Entanglement of qubit pairs within a chain of $N=5$ trapped ions. 
(a) Comparison of theoretical entangled state fidelity for a constant pulse (black) versus a nine-segment pulse (red) when the gate is performed on ion pair 1\&2. The black arrow indicates the optimal detuning for the constant pulse. The right panels show measured parity oscillation for the gate detuning indicated by the red arrow along with the segmented pulse pattern used at this detuning. 
(b) Same as (a), except the gate is performed on ion pair 2\&3, with the gate detuning indicated by the blue arrow. 
(c) Phase space trajectories (arbitrary units) for the solution on ion pair 1\&2 in at the detuning indicated by the red arrow in (a).}
\label{fig:segmentCompare}
\end{figure}
}

\newcommand{\FigureThree}{
\begin{figure}[btp!]
\includegraphics*[width=0.7\columnwidth]{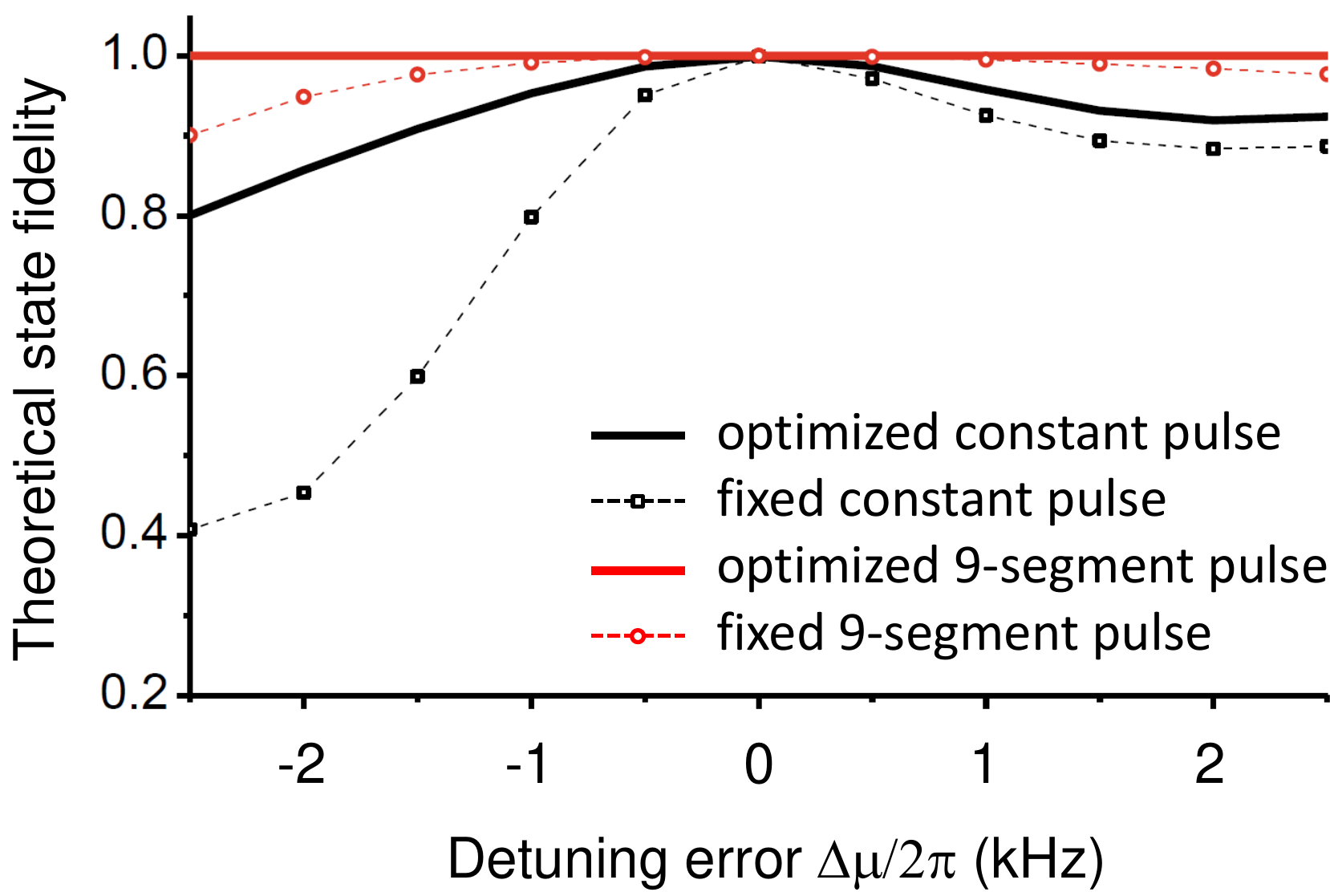}
\caption{Theoretical entangled state fidelity as a function of detuning error $\Delta \mu$. The black (red) line corresponds to a constant (nine-segment) pulse shape on ion pair 1\&2 in a five-ion chain at the detunings indicated by the black (red) arrows in Fig. \ref{fig:segmentCompare}a.  The multi-segment approach is less sensitive to detuning (or trap frequency) fluctuations.}
\label{fig:FidelityStability}
\end{figure}
}

\newcommand{\FigureFour}{
\begin{figure}[htbp!]
\includegraphics*[width=\columnwidth]{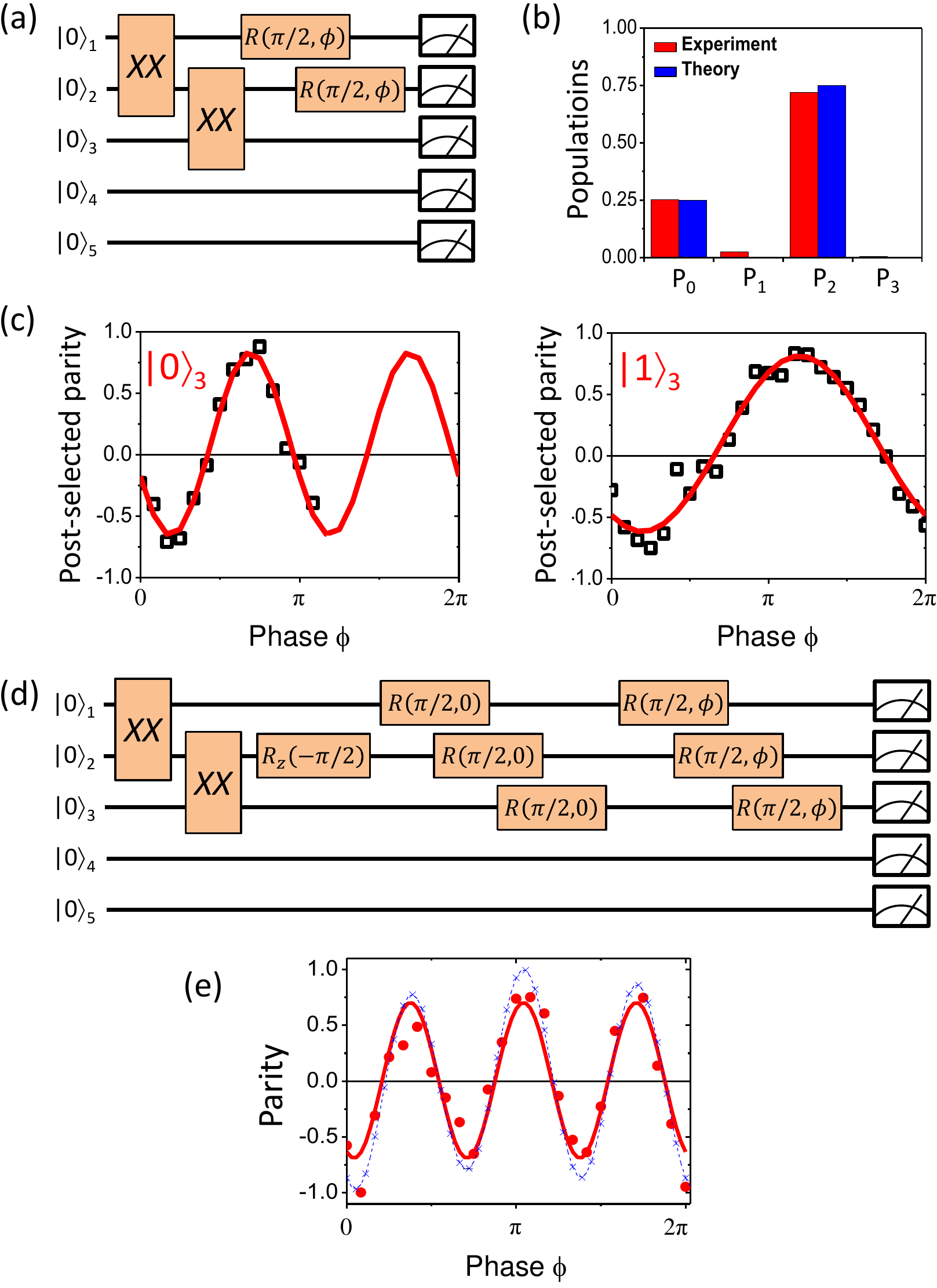}
\caption{Programmable quantum operations to create tripartite entanglement. 
(a) Circuit for concatenated \textit{XX} gates between ions 1\&2 and 2\&3 and $\pi/2$ analysis rotations of ions 1\&2 with phase $\phi$. 
(b) Measured population after \textit{XX} gates on ions 1\&2 and 2\&3, where P$_N$ denotes the probability of finding $N$ ions in the $\ket{1}$ state. 
(c) Parity oscillations of ions 1\&2 with the phase $\phi$ of the $\pi/2$ analysis rotations, after post-selecting the state of the third ion, with periods $\pi$ (left) and  $2\pi$ (right) for the two states $\ket{0}_3$ and $\ket{1}_3$, respectively (see Eq. (\ref{eq:phiState})). 
(d) Schematic for creating a GHZ ``cat" state using two \textit{XX} gates on ions 1\&2 and 2\&3 as before, with additional individual qubit rotations, followed by $\pi/2$ analysis rotations of all three ions with phase $\phi$. 
(e) Three-ion parity oscillation with phase $\phi$ of the $\pi/2$ analysis rotations. The red solid line is fit to the data with period $2\pi/3$, while the blue dashed line is the expected signal assuming a perfect cat state with known systematic measurement errors. }
\label{fig:tripartiteEntanglement}
\end{figure}}

\begin{document}

\newcommand{\ket}[1]{|#1\rangle}
\newcommand{\bra}[1]{\langle #1|}
\newcommand{\Yb}{$^{171}{\rm{Yb}}^{+} $}
\newcommand{\up}{\uparrow}
\newcommand{\dn}{\downarrow}
\newcommand{\upr}{\uparrow\rangle}
\newcommand{\dnr}{\downarrow\rangle}
\newcommand{\sx}{\hat{\sigma}_{x}}
\newcommand{\sy}{\hat{\sigma}_{y}}
\newcommand{\sz}{\hat{\sigma}_{z}}

\title{Optimal quantum control of multi-mode couplings \\ between trapped ion qubits for scalable entanglement}

\author{T. Choi}
\author{S. Debnath}
\author{T. A. Manning}
\author{C. Figgatt}
\affiliation{Joint Quantum Institute, University of Maryland Department of Physics and \\
                    National Institute of Standards and Technology, College Park, MD  20742}
\author{Z.-X. Gong}
\affiliation{Joint Quantum Institute, University of Maryland Department of Physics and \\
                    National Institute of Standards and Technology, College Park, MD  20742}
\affiliation{Department of Physics, University of Michigan, Ann Arbor, MI 48109, USA}
\author{L.-M. Duan}
\affiliation{Department of Physics, University of Michigan, Ann Arbor, MI 48109, USA}
\author{C. Monroe}
\affiliation{Joint Quantum Institute, University of Maryland Department of Physics and \\
                    National Institute of Standards and Technology, College Park, MD  20742}
\date{\today}

\begin{abstract}
 We demonstrate high fidelity entangling quantum gates within a chain of five trapped ion qubits by optimally shaping optical fields that couple to multiple collective modes of motion.  We individually address qubits with segmented optical pulses to  construct multipartite entangled states in a programmable way. This approach enables both high fidelity and fast quantum gates that can be scaled to larger qubit registers for quantum computation and simulation.
\end{abstract}

\maketitle
Trapped atomic ion crystals are the leading architecture for quantum information processing, with their unsurpassed level of qubit coherence and near perfect initialization and detection efficiency \cite{Blatt08,Monroe13}. Moreover, trapped ion qubits can be controllably entangled through their Coulomb-coupled motion by applying external fields that provide a qubit state-dependent force \cite{Cirac95,Molmer99,Solano99,Milburn00}. 
However, scaling to large numbers 
of ions $N$ within a single crystal is complicated by the many collective modes of motion, which can cause gate errors from mode crosstalk.  Such errors can be mitigated by coupling to a single motional mode, at a cost of significantly slowing the gate operation.  The gate time $\tau_g$ must generally be much longer than the inverse of the frequency splitting of the motional modes, which for axial motion in a linear chain implies $\tau_g \gg 1/\omega_{z} > N^{0.86}/\omega_{x} $, where $\omega_{z} $ and  $\omega_{x} $ are the center-of-mass axial and transverse mode frequencies \cite{Schiffer93}.  For gates using transverse motion in a linear chain \cite{Zhu06}, we find $\tau_g \gg \omega_{x}/\omega_{z}^2 > N^{1.72}/\omega_{x} $.  In either case, the slowdown with qubit number $N$ can severely limit the practical size of trapped ion qubit crystals.  

In this letter, we circumvent this scaling problem by applying qubit state-dependent optical forces that simultaneously couple to multiple modes of motion.  We address subsets of ions immersed in a five-ion linear crystal and engineer laser pulse shapes to entangle pairs of ions with high fidelity while suppressing mode crosstalk and maintaining short gate times \cite{Zhu061,Zhu06}. 
The pre-calculated pulse shapes optimize target gate fidelity, achieving unity for sufficiently complex pulses.
In the experiment, we concatenate these shaped gates to entangle multiple pairs of qubits and directly measure multi-qubit entanglement in the crystal. Extensions of this approach can be scaled to larger ion chains and also incorporate higher levels of pulse shaping to reduce sensitivity to particular experimental errors and drifts \cite{Kirchmair09,Hayes12,Tomita10}.

In the experiment, five \Yb~ions are confined in a three-layer linear rf trap \cite{Hensinger06} with transverse center-of-mass (CM) frequency ranging from $\omega_x/2\pi=2.5-4.5$~MHz and axial CM frequency $\omega_z/2\pi=310-550$~kHz, with a minimal ion separation of  $\sim 5$~$\mu$m.
Each qubit is represented by the $^2S_{1/2}$ hyperfine ``clock" states within \Yb, denoted by 
$\ket{0}$ and $\ket{1}$ and having a splitting of $\omega_0/2\pi=12.642821$~GHz \cite{Olmschenk07}. We initialize each qubit by optically pumping to state $\ket{0}$ using laser light resonant with the $^2S_{1/2}~\leftrightarrow~^2P_{1/2}$ transition near $369.5$~nm.  

We then coherently manipulate the qubits with a mode-locked laser at $355$~nm whose frequency comb beat notes drive stimulated Raman transitions between the qubit states and produce qubit state-dependent forces \cite{Hayes10, Campbell10}.  
The Raman laser is split into two beams, one illuminating the entire chain and the other focused to a waist of $\sim 3.5$~$\mu$m for addressing any subset of adjacent ion pairs in the chain, with a wavevector difference $\boldsymbol{\Delta k}$ aligned along the $x$-direction of transverse motion. We finally measure the state of each qubit by applying resonant laser light near $369.5$~nm that results in state-dependent fluorescence \cite{Olmschenk07} that is imaged onto a multi-channel photo-multiplier tube (PMT) for individual qubit state detection.  We repeat each experiment at least $300$ times and extract state populations by fitting to previously measured fluorescence histograms \cite{Acton06}. 

When a constant state-dependent force is applied to the ion qubits, the multiple incommensurate modes generally remain entangled with the qubits following the interaction, thereby degrading the quantum gate fidelity.  However, more complex optical pulses can be created that satisfy a set of constraints for disentangling every mode of motion following the gate.  This optimal control problem involves engineering a sufficiently complex laser pulse that maximizes or even achieves unit fidelity.

The qubit state-dependent optical force is applied by generating bichromatic beat notes near the upper and lower motional sideband frequencies at $\omega_0\pm\mu$, where the detuning $\mu$ is in the neighborhood of the motional mode frequencies.
 Using the rotating wave approximation in the Lamb-Dicke limit, the evolution operator of the dipole interaction Hamiltonian becomes \cite{Zhu03,Zhu061,Kim09}

\begin{equation}
\hat{U}(\tau) = \mathrm{exp}\left[\sum_{i}^{}\hat{\phi}_i(\tau)\sx^{(i)}+i\sum_{i,j}^{}\chi_{i,j}(\tau)\sx^{(i)}\sx^{(j)}\right].
\label{eq:eq1}
\end{equation}
The first term corresponds to the qubit-motion coupling on ion \textit{i}, where 
$\hat{\phi}_i(\tau)=\sum_{m}^{}\left[\alpha_{i,m}(\tau)\hat{a}^\dagger_m- {\alpha_{i,m}}^*(\tau)\hat{a}_m \right]$,
$\hat{a}^\dagger_m (\hat{a}_m)$ is the raising (lowering) operator of mode $m$, and 
$\sx^{(i)}$ is the Pauli-X operator of the ith qubit, where we define the x-axis of the qubit Bloch sphere according to the phase of the bichromatic beatnotes \cite{LeeReview}.
This is a state-dependent displacement of the ion \textit{i} such that the $\ket{0} \pm \ket{1}$ states follow the trajectories $\pm\alpha_{i,m}(\tau)$ in phase space of the $m^{th}$ motional mode according to \cite{Zhu061}
\begin{equation}
\alpha_{i,m}(\tau)=i\eta_{i,m}\int_0^\tau\Omega_i(t) \mathrm{sin}(\mu t)e^{i\omega_m t}dt.
\label{eq:eq2}
\end{equation}
Here, $\eta_{i,m}=b_{i,m}\cdot\Delta k \sqrt[]{\hbar/2M\omega_m}$ is the Lamb-Dicke parameter, $b_{i,m}$ is the normal mode transformation matrix for ion $i$ and mode $m$ \cite{James98}, $\omega_m$ is the frequency of the $m^{th}$ motional mode, and $M$ is the mass of a single \Yb~ion. 
The second term of Eq. (\ref{eq:eq1}) describes the entangling interaction between qubits $i$ and $j$, with \cite{Zhu061}
\begin{equation}
\begin{split}
\chi_{i,j}(\tau) =2 & \sum_{m} \eta_{i,m} \eta_{j,m} \int_0^\tau\int_0^{t'} \Omega_i(t) \Omega_j(t') \\
& \times \mathrm{sin}(\mu t) \mathrm{sin}(\mu t')\mathrm{sin}[\omega_m(t'-t)]dtdt'.
\end{split}
\label{eq:eq3}
\end{equation}
In Eqs. (\ref{eq:eq2}-\ref{eq:eq3}), the time-dependent Rabi frequency $\Omega_i(t)$ on the $i$th ion is used as a control parameter for optimization of the gate and is assumed to be real without loss of generality.  (We could alternatively vary the detuning 
$\mu$ over time for control \cite{Korenblit12}.) 

In order to perform an entangling $XX$ gate on two ions $a$ and $b$ in a chain of $N$ ions, 
we apply identical state-dependent forces to just these target ions 
and realize $\hat{U}(\tau_g)=\mathrm{exp}[i\pi\sx^{(a)}\sx^{(b)}/4]$.  This requires 
$\chi_{a,b}(\tau_g)=\pi/4$ along with the $2N$ conditions 
$\alpha_{a,m}(\tau_g)=0$ so that the phase space trajectories of all $N$ motional modes return to their origin and disentangle the qubits from their motion \cite{Molmer99,Solano99,Milburn00}. These constraints can be satisfied by evenly partitioning the pulse shape $\Omega_{a}(t)=\Omega_{b}(t)$ into $2N+1$ segments \cite{Zhu06,Zhu061}, reducing the problem to a system of linear equations with a guaranteed solution. The detuning and gate duration become independent parameters so that in principle, the gate can be performed with unit fidelity at any detuning $\mu \ne \omega_m$ and any gate speed on any two ions in a chain, given sufficient optical power. 

When the gate is faster than the trap frequencies \cite{GarciaRipoll03, DuanFast}, the motion of the target ions is excited and stopped faster than the response time of the chain. In this case, the motion is better described using the basis set of ``local modes" involving only the two target ions, thereby reducing the control problem to $2N+1=5$ equations regardless of the total number of ions in the chain 
\cite{Zhu061, DuanFast}.  In the experiment, the minimum achievable gate time of $\tau_g  \sim $20 $\mathrm{\mu}$s is considerably longer than the trap period of $2\pi/\omega_x < 1$~$\mu$s, implying that all $2N+1$ control parameters are required.  However, a judicious choice of detuning can often reduce the number of parameters required to achieve near-unit gate fidelities \cite{Zhu061,Zhu06}. 

\FigureOne

Figure \ref{fig:fullControl}a shows theoretical and measured fidelity of the Bell state 
$\hat{U}(\tau_g)\ket{00}=\ket{00}+i\ket{11}$ for both a simple constant pulse and a five segment pulse on a two-ion chain, as a function of detuning $\mu$ for a fixed gate time $\tau_g=104$~$\mu$s.  
For two ions, the five segments provide full control ($2N+1=5$), meaning that a pulse shape can be calculated at each detuning that should yield unit fidelity.  As seen in Fig. \ref{fig:fullControl}a, a constant pulse can be optimized to achieve high fidelity, but only at detunings 
$\mu$ whose frequency difference from the two modes is commensurate \cite{Kim09}, which in this case has many solutions spaced by 
$1/\tau_g$. The observed fidelity of the constant pulse follows theory, with uniformly lower fidelities consistent with known errors in the system.
On the other hand, high fidelities of the five segment pulse are observed over a wide range of detunings for the same gate time, with the details of a particular pulse sequence shown in Fig.  \ref{fig:fullControl}b-c.  
We measure the fidelity by first observing the populations of the $\ket{00}$ and $\ket{11}$ states, then extracting their coherence by repeating the experiment with an additional global $\pi/2$ analysis rotation $R(\pi/2,\phi)$ 
and measuring the contrast in qubit parity as the phase $\phi$ is scanned \cite{Sackett00}. 

\FigureTwo

When the number of ions in a chain increases to $N>2$, it becomes difficult to find detunings $\mu$ of a constant pulse whose difference frequencies $\mu - \omega_m$ from all modes are nearly commensurate, without significantly slowing the gate. 
Figure \ref{fig:segmentCompare} shows the state fidelity for a constant pulse versus a nine-segment pulse for entangling adjacent ion pairs $1\&2$ or $2\&3$ within a five ion chain, with gate time $\tau_g = 190~\mu$s. We find significant improvement over a wide range of detunings when using more segments, even though fewer than $2N+1=11$ control parameters are utilized. Using nine-segment pulses, we achieve state fidelities over 95(2)\% for ion pairs 1\&2 and 2\&3 at the detunings indicated by the red and blue arrows in Fig. \ref{fig:segmentCompare}a-b. In this overconstrained case, the calculation becomes an optimization problem, where more weight is given to the closing of more influential phase space trajectories (Fig. \ref{fig:segmentCompare}c).  

\FigureThree

A further advantage of using multi-segment pulses is their relative insensitivity to fluctuations in detuning $\mu$ and trap frequency $\omega_m$. Such drifts cause errors because the pulse shape is no longer optimal.  However, multi-segment pulses can significantly mitigate this error \cite{Hayes12}, admitting solutions that do not change rapidly with detuning. As seen in Fig. \ref{fig:FidelityStability}, a constant pulse is expected to degrade the fidelity by $\sim15\%$ for a $1$~kHz drift in detuning, which is consistent with the measured state fidelity of 82(3)\%. However, the nine-segment pulse is expected to degrade the fidelity by only $1\%$ for the same drift, which compares to the observed fidelity of $95(2)\%$.

\FigureFour

To demonstrate pulse-shaped gates on subsets of qubits in a linear crystal, we produce tripartite entangled states by concatenating two $XX$ gates in a five ion chain (see Fig.  \ref{fig:tripartiteEntanglement}a). We adiabatically shuttle the ions across the fixed laser beams in order to address nearest neighbor pairs of the three target ions and ideally create a GHZ-type state
\begin{equation}
\ket{0 0 0} \rightarrow \ket{0 0 0} + i \ket{1 1 0} + i \ket{0 1 1} - \ket{1 0 1}.
\label{eq:phiState}
\end{equation}
The measured state populations are consistent with the above state, as shown in Fig.  \ref{fig:tripartiteEntanglement}b. 

In order to measure the coherences of the three-qubit subsystem, we apply  analysis rotations $R(\pi/2,\phi)$ to any two of the three qubits, then measure their parity as before. (The individual rotations are accomplished by adiabatically weakening the axial trap confinement and shuttling the ions so that the focused Raman laser beam addresses just the target ion.)
As the phase $\phi$ of the analysis rotations is scanned, the parity should oscillate with period $\pi$ or $2\pi$ when the third ion is post-selected to be in state $\ket{0}$ or $\ket{1}$, respectively, as seen in Fig.  \ref{fig:tripartiteEntanglement}c for one of the pairs. 
By measuring the contrasts of the two parity curves for each of the three possible pairs conditioned upon the measured value of the third, we obtain the six coherences of the final state.
Combined with the state populations (Fig. \ref{fig:tripartiteEntanglement}b), we calculate a quantum state fidelity of 79(4)\% with respect to Eq. (\ref{eq:phiState}). This level of fidelity is consistent with the compounded $XX$ gate fidelities ($\sim 95\%$ each) and the discrimination efficiency ($\sim 93\%$) for post-selection of the third qubit. 

To prove genuine tripartite entanglement within the five ion chain, we use single qubit rotations to transform the state given by Eq. \ref{eq:phiState} into a GHZ ``cat'' state $\ket{0 0 0} + i \ket{1 1 1}$ \cite{Dur99}.  As shown in the circuit of 
Fig.  \ref{fig:tripartiteEntanglement}d, this is achieved by applying a Z-rotation operation 
$R_z(-\pi/2) = R(-\pi/2,0)R(\pi/2,\pi/2)R(\pi/2,0)$ to the middle ion only followed by 
$R(\pi/2,0)$ rotations to all three ions.
We finally measure the parity of all three qubits while scanning the phases of subsequent $R(\pi/2,\phi)$ analysis pulses, and the oscillation with period $2\pi/3$ with a contrast of over 70\% (Fig. \ref{fig:tripartiteEntanglement}e) verifies genuine tripartite entanglement \cite{Sackett00}. 
This is a conservative lower limit to the entanglement fidelity, given known errors and crosstalk in the rotations and the measurement process. The simulated blue dashed curve in the same figure depicts what we expect to measure given our known errors and assuming a perfect initial state. 


We have shown how a single control parameter can be used to mitigate multi-mode couplings between a collection of qubit, but this approach can be expanded to include additional parameters, such as spectral and spatial addressing of each qubit \cite{Korenblit12, DuanPRA13}.  This could allow for the efficient implementation of more complicated quantum circuits, such as Toffoli \cite{Toffoli} and other gates involving more than two qubits, or global operations for quantum simulations of particular Hamiltonian models \cite{Feynman}.  The optimal quantum control we demonstrate here could apply to any quantum information and simulation architectures that entangle subsets of qubits through a bosonic quantum bus having multi-mode components, such as cavity QED \cite{CQED} and superconducting circuits \cite{SCReview}. 

This work is supported by the U.S. Army Research Office (ARO) Award W911NF0710576 with funds from the DARPA Optical Lattice Emulator Program, ARO award W911NF0410234 with funds from the IARPA MQCO Program, ARO MURI award W911NF0910406, and the NSF Physics Frontier Center at JQI.

\bibliographystyle{prsty}

\end{document}